\documentclass{article}

\usepackage{amsmath}
\usepackage{amssymb}
\usepackage{times}
\usepackage{bm}
\usepackage{natbib}
\usepackage{bm}
\usepackage[pdftex]{graphicx}
\usepackage{multirow}
\usepackage{amsthm}
\usepackage{amsmath}

\usepackage{appendix}

\usepackage[plain,noend]{algorithm2e}

\makeatletter
\renewcommand{\algocf@captiontext}[2]{#1\algocf@typo. \AlCapFnt{}#2} 
\def\@algocf@capt@plain{top}
\renewcommand{\algocf@makecaption}[2]{%
  \addtolength{\hsize}{\algomargin}%
  \sbox\@tempboxa{\algocf@captiontext{#1}{#2}}%
  \ifdim\wd\@tempboxa >\hsize
    \hskip .5\algomargin%
    \parbox[t]{\hsize}{\algocf@captiontext{#1}{#2}}
  \else%
    \global\@minipagefalse%
    \hbox to\hsize{\box\@tempboxa}
  \fi%
  \addtolength{\hsize}{-\algomargin}%
}
\makeatother

\newtheorem{prop}{Proposition} 
\newtheorem{lmma}{Lemma} 
\newtheorem{thm}{Theorem}

\begin{document}


\title{Bayesian Local Extrema Splines}

\author{M. W. WHEELER, D. B. Dunson, A. H. HERRING}

	\maketitle
\abstract{ We consider the problem of shape restricted nonparametric regression on a closed 
set $\mathcal{X} \subset \mathbb{R},$ where it is reasonable to assume the function 
has no more than $H$ local extrema interior to $\mathcal{X}.$
Following a Bayesian approach we develop a nonparametric prior over a 
novel class of local extrema splines. This approach is shown to be consistent when modeling
any continuously differentiable function within the class of functions considered, and 
is used to develop methods for hypothesis testing on the 
shape of the curve. Sampling algorithms are developed, and the 
method is applied in simulation studies and data examples where
the shape of the curve is of interest.}

\section{Introduction}\label{sec:one}
This paper considers Bayesian modeling of an unknown function  $f_0:\mathcal{X} \rightarrow \mathbb{R},$ 
where it is known that $f_0$ has at most $H$ local extrema, or change points, interior to $\mathcal{X}$,
and one wishes to estimate the function subject to constraints or test the hypothesis the function has a specific shape. For example,
one may wish to consider a monotone function as compared to a function having an  `N' shape. 
We propose a novel spline construction that allows for nonparametric estimation of shape 
constrained functions having at most $H$ change points. The approach places a novel prior 
over a knot set that is dense in $\mathcal{X}$ while developing Markov chain Monte Carlo
algorithms to sample between models. The method allows for nonparametric hypothesis testing of 
different shapes within the class of functions considered using Bayes factors.
 
The shape constrained regression literature focuses primarily on functions that 
are monotone, convex, or have a single minimum, that is, cases with $H \leq 1$. 
\citet{ramgopal1993}, \citet{lavine1995}, and \citet{bornkamp2009} consider
priors over cumulative distribution functions used to model 
monotone curves. Alternatively, \citet{holmes2003}, \citet{neelon2004}, \citet{meyer2008}, and \citet{shively2009}    
develop spline based approaches for monotone functions. 
\citet{hans2005} design a prior for umbrella-shaped functions, while \citet{shively2011} propose
methods for fixed and free knot splines that model continuous segments having a single unknown  change 
point. 

Extending these approaches to broader shape constraints is not straightforward computationally.  For example,
to obtain $H=3$ change points, one could define a prior over B-spline bases 
(\citet{deBoor2001}, page 87) having four monotone segments alternating between increasing and decreasing. 
Even for a moderate number of pre-specified knots and a known number of change points, allowing for uncertainty in 
the locations of the change points leads to a daunting computational problem.  For example, Bayesian computation
via Markov chain Monte Carlo is subject to slow mixing and convergence rates in alternating between updating 
the spline coefficients conditionally on the change points and vise versa.  It is not clear how to devise algorithms that 
can efficiently update change points and coefficients simultaneously.  These difficulties are 
compounded by allowing for the possibility that some of the change points should be removed, which is commonly 
the situation in applications.   By defining a new spline basis based on the number of change points, we 
bypass these issues.

Also, little work has been done on nonparametric Bayesian testing of curve shapes.
Recently, \citet{salomond2014} and \citet{scott2015} consider Bayesian nonparametric 
testing for monotonic versus an unspecified nonparametric alternative, but do not 
consider shapes beyond monotonicity. Our approach is different because it allows 
for testing of all shapes, where shape is defined as the type and sequence of extrema. 
For example, one can use this approach to test for an
umbrella shape verses an `N' shaped curve and use the same procedure 
to test the umbrella shape against monotone alternatives.   

We propose a fundamentally new approach to incorporating shape constraints based on splines that are carefully constructed to induce curves having a particular number of extrema.  This is similar in spirit to the I-spline construction of \citet{ramsay1988} or the C-spline construction for convex splines 
\citep{meyer2008, meyer2011}, which both create a spline construction based upon the derivative of the spline.
Our spline construction, when paired with positivity constraints on the spline coefficients, enforces shape 
restrictions on the curve of interest by limiting the number of change points. 

Another key aspect of our approach is that we place a prior over a model space that can grow to a 
countably dense set of knots. This bypasses  the sensitivity to choice of the number of knots, 
while facilitating computation and theory on consistency. 
In particular, we propose a prior over nested model spaces where the location of the knots is known for each model. 
This allows for a straightforward reversible jump 
Markov chain Monte Carlo algorithm \citep{green1995} based upon \citet{godsill2001}. This is different from much of the previous Bayes 
literature allowing unknown numbers of knots \citep{biller2000,dimatteo2001}. In these methods, 
the knot locations are unknown, and the reversible jump Markov chain Monte Carlo 
proposal must propose a knot to add or delete as well as its location.  Such algorithms are notoriously inefficient.

\section{Model}
\subsection{Local Extrema Spline Construction}
Let $\mathcal{F}^{H}$ be a set of functions defined on the closed set $\mathcal{X} \subset \mathbb{R},$ 
such that for $f_0\in \mathcal{F}^{H},$  $f_0$ is continuously differentiable and has 
$H$ or fewer local extrema interior to $\mathcal{X}.$
Such functions can be modeled using B-spline approximations of the form 
\begin{align}
	f(x)  = \sum_{k=1}^{K+j-1} \beta_k B_{(j,k)}(x), \label{BX:splineapprox}
\end{align}
Here, $\beta_k$ is a scalar coefficient, and $B_{(j,k)}(x)$ is a B-Spline 
function of order $j$ defined on the knot set $\mathbf{\mathcal{T}} = \{\tau_k\}_{k=1}^{K},$ 
$\tau_1 \leq \tau_2 \leq \ldots \leq \tau_K,$ which includes end knots.    
\citet{deBoor2001}, page 145, showed that for any knot set there exists 
spline approximations such that $||f-f_0||_{\infty} \leq \Delta||f_0||_{\infty},$
where $\Delta$ is  the  maximum difference between adjacent knots.  
Though this construction can be used to model $f_0$ with arbitrary accuracy,
it does not guarantee the approximating function $f$ is itself in $\mathcal{F}^{H}.$

We force $f \in \mathcal{F}^{H}$  to have at most 
$H$ local extrema by defining a new spline basis
\begin{align}
	B_{(j,k)}^{\ast}(x) = M \int_{-\infty}^{x} \bigg\{ \prod_{h=1}^H (\xi - \alpha_h) \bigg\} B_{(j,k)}(\xi) d\xi, \label{BX:splineconstruction}
\end{align}
where, as above, $B_{(j,k)}(x)$ is a B-spline that is
constructed using the knot set $\mathcal{T},$  $\{\alpha_1,\ldots,\alpha_h\}$ are distinct change points
and the scalar $M$ is a fixed integer. Letting
$B_{(j,0)}^{\ast}(x) = 1,$  if $\beta_k \geq 0$, for all $k \geq 1,$
then any linear combination of local extrema spline basis functions for any 
distinct values of $\alpha_1,\ldots,\alpha_H$ in (\ref{BX:splineconstruction})
will be in $\mathcal{F}^{H}.$  

\begin{prop}
Letting $f(x) = \sum_{k=0}^{K+j-1} \beta_k B^{\ast}_{(j,k)}(x)$ for any $K \geq 1$
with $M \in \{-1,1\},$ $j \geq 1,$ and $\beta_k \geq 0$ for all $k \geq 1,$ then $f \in \mathcal{F}^{H}.$ 
\end{prop}

This result follows from the constraint on the $\beta_k$ coefficients.  By forcing $\beta_k \geq 0$ for $k \geq 1$, 
the sign of the derivative is controlled by the polynomial $M\prod_{h=1}^H(x-\alpha_h),$ 
which forces a maximum of  $H$ local extrema located at the change points $\{\alpha_1,\ldots,\alpha_H\}$. 
 When $\beta_k =\ldots = \beta_{k+j} = 0$ and $\alpha_h \in [\tau_{k+j},\tau_{k+j+1}],$ $\alpha_h$ does not define unique extrema.  In this case, there is a flat region
and multiple configurations of the change point parameters can result in the same curve.
Otherwise, the extrema are uniquely defined for all $\alpha_h \in \mathcal{X},$ and
fewer than $H$ extrema can be considered if $\alpha_h \notin \mathcal{X}.$

\begin{thm}
For any $f_0 \in \mathcal{F}^{H}$ and $\epsilon > 0$ there exists a knot set $\mathcal{T}$ and a local extrema spline $f^{LX}$ defined
on this knot set such that
\begin{align*}
	\| f_0 - f^{LX}\|_\infty < \epsilon.
\end{align*}
\end{thm}

The flexibility of local extrema splines is attributable to the B-splines used in their construction.  
The proof of this theorem assumes that $M$ can be chosen to be positive or negative, which allows
all functions in $\mathcal{F}^{H}$ to be approximated.   
If $M$ is fixed, then any function with $H-1$ extrema can be modeled. For functions with exactly $H$ extrema, one is limited to modeling functions that are either initially increasing or initially decreasing, and this depends on the sign of $M$.

\textit{Remark:} Though the polynomial 
weighting does not affect the ability of the local extrema spline to model arbitrary functions
in $\mathcal{F}^{H},$ it does impact the magnitude of the spline, that is, $\sup_{x\in \mathcal{X}} |B^{\ast}_{(j,k)} (x)|,$ 
which may cause difficulty in the prior specification.
To minimize this effect it is often beneficial to construct the splines on
the interval $(-0{\cdot}5,0{\cdot}5).$ Additionally, it is often beneficial to multiply M by a fixed constant
to aid in prior specification. 
 
\subsection{Infill Process Prior}
Bayesian methods for automatic knot selection \citep{biller2000,dimatteo2001}
commonly define  priors over the number and location of knots.  Using free knots presents computational 
challenges while fixed knots are too inflexible; we address this by defining a prior over a branching process 
where the children of each generation represent knot locations that are binary infills of the previous 
generation.  This defines a nested set of spline models such that successive generations
produce knot sets  that are  arbitrarily close. 

To make these ideas explicit, define  $\mathcal{T}_N = \{{a}/{2^{N+1}}: a = 1,3,\ldots,2^{N+1}-1\}$
with $N \in \{0,1,2,3,\ldots\}.$ Assume $\mathcal{X} = [0,1]$ for the sake of exposition,
and consider an infinite complete binary tree. 
In this tree, each node at a given depth $N$ is uniquely labeled using 
an element from $\mathcal{T}_N.$ Given the node's label is 
${a}/{2^{N+1}},$  its  children are labeled 
${(2a-1)}/{2^{N+2}}$ and ${(2a+1)}/{2^{N+2}}.$  
For example,  the node labeled ${3}/{8}$ at $N=2$ has children labeled ${5}/{16}$
and ${7}/{16},$ and the root node labeled ${1}/{2}$ has children labeled ${1}/{4}$
and ${3}/{4}.$ 

We induce a prior on the set of local extrema spline basis functions through a branching process over this tree. The process starts at the root node $N=0$ where the generation of children occurs via
two independent Bernoulli experiments having probability of success $\zeta.$ 
On each success, a child is generated, and its label is added to the knot set. This process repeats until it dies out.  If $\zeta < 0{\cdot}5$, the probability of extinction is $1$ 
(\citet{feller1974}, page 297).  To favor parsimony in the tree, we define the 
probability of success for a node at a given depth $N$ to be $0{\cdot}5^{N+1},$ which 
decreases the probability of adding a new node the larger the tree becomes. 
The tree $\mathcal{M}$ generated from this process corresponds to a knot set  $\mathcal{T}_{\mathcal{M}}$. 
We complete the knot set by adding end knots $\{0,1\}.$
 
Letting $K = |\mathcal{T}_{\mathcal{M}}|$ be the number of knots for tree $\mathcal{M}$ including end knots, 
there are $K+j-1$ basis functions.
Letting $\beta_k \in \beta_{\mathcal{M}}$ denote the coefficient on $B^{\ast}_{(j,k)}(x),$ we choose the prior: 
\begin{align}
	p(\beta_k | \mathcal{M}) =  \pi1_{(\beta_k = 0)} + (1-\pi) \text{Exp}(\beta_k; \lambda),\quad 1 \leq k \leq K+j-1, \label{PRI:1} 
\end{align}
where $\text{Exp}(\lambda)$ is an exponential distribution with rate parameter $\lambda$, $\pi$  is the prior probability of $\beta_k =  0$, 
and the $\beta_k$ are drawn independently conditionally on $\mathcal{M}, \pi, \lambda$.  
For the intercept, we let $\beta_0 \sim N(0,c)$, and we allow for greater adaptivity to the data through hyperpriors, $\pi \sim \mbox{Be}(\nu,\omega)$ and $\lambda \sim \mbox{Ga}(\delta,\kappa)1(\lambda > \epsilon),$ which
is a truncated gamma distribution, truncated slightly above zero to guarantee posterior consistency. In
practice, this value is set to $1e-5$ making the prior indistinguishable from the Gamma distribution. 

To allow uncertainty in locations of the change points, we choose the prior 
\begin{align}
	p(\alpha) = \prod_{h=1}^{H}\text{TN}\{\alpha_h; (b-a)/2,1,a,b\} \label{PRI:3}
\end{align}
where  $\text{TN}\{(b-a)/2,1,a,b\}$ is truncated normal with mean $(b-a)/2,$ variance $1$, and is
truncated below by $a$ and above by $b$ with $\mathcal{X} \subset [a,b].$  If $\alpha_h \leq \inf \mathcal{X}$
or $\alpha_h \geq \sup \mathcal{X}$, then the change point is removed.  
We assume $M$ is pre-specified corresponding to prior knowledge of whether the function is initially 
increasing or decreasing, though generalizations to place a prior on $M$, for example a Bernoulli prior on $M,$ are straightforward.  

\textit{Remark:} The prior for the change point parameters is defined such that $\mathcal{X} \subset [a,b].$ 
When a change point is placed outside of $\mathcal{X}$, this allow for the derivative of $f$ to be non-zero at 
$\inf \mathcal{X}$ or  $\sup \mathcal{X}.$  
In practice, results are insensitive to the choice of $a$ and $b.$
In what follows, we chose $a=\inf (\mathcal{X}) - \Delta$ and $b = \sup (\mathcal{X}) + \Delta$ where
$\Delta = \{\sup (\mathcal{X}) - \inf (\mathcal{X})\}/2.$

\subsection{Prior Properties}
Define $\mathcal{F}^{H+}$ as the space of 
continuously differentiable functions with $H$ or fewer local extrema, such that, for all 
$f_0 \in \mathcal{F}^{H+}$  
having exactly $H$ extrema, the first extrema from the left is a maximum, and, for all functions in
$f_0 \in \mathcal{F}^{H+}$   having less than $H$ extrema, the function is also in  $\mathcal{F}^{H-1}.$ 
Conversely, define $\mathcal{F}^{H-}$  as the set of continuously differentiable functions with $H$ or fewer local extrema,
such that, for all functions having exactly $H$ extrema the first from the left is a minimum, 
and for all functions $f_0 \in \mathcal{F}^{H-}$  having less than $H$ extrema they  are also in $\mathcal{F}^{H-1}.$  The prior places positivity in $\epsilon-$neighborhoods of any $f_0$ in $\mathcal{F}^{H-}$ or $\mathcal{F}^{H+}$ depending on the sign of $M$.  

\begin{lmma}
	Letting $f^{LX}$ be a randomly generated local extrema spline from the prior defined in $\S 2{\cdot}2$ 
        for all $f_0\in \mathcal{F}^{H-1}:$   
	\begin{align*}
		   \mbox{pr}(||f^{LX} - f_0||_{\infty} < \epsilon) > 0.
	\end{align*}
This holds for all $f_0 \in \mathcal{F}^{H+}$ if $H$ is odd and $M<0$ or $H$ is even and $M>0$.  Otherwise, if $H$ is even and $M>0$ or $H$ is odd and $M<0$, this holds for all $f_0 \in \mathcal{F}^{H-}$.   
\end{lmma} 
 
Using this result we can show posterior consistency. 
Assume $Y = (y_1,\ldots,y_n)^T$ are observed at locations 
$(x_1,\ldots,x_n)$ such that $y_i \sim \text{N}\{f_0(x_i),\sigma_0^2\}.$ Following 
\citet{choi2007}, assume that the design points
are drawn independent and identically distributed from some probability distribution $Q$ on the interval $\mathcal{X},$ or observed
using a fixed design such that $\max(|x_i - x_{i+1}|) < (K_1 n)^{-1}$ where $0 < K_1 < 1$ and  
$i < n.$ Also,  define the neighborhoods $W_{\epsilon,n} = \{(f,\sigma): \int |f(x)-f_0(x)| dQ_n(x) < \epsilon, |\sigma/\sigma_0 - 1 | < \epsilon\}$
and $U_{\epsilon} = \{(f,\sigma): d_Q(f,f_0) < \epsilon, |\sigma/\sigma_0 - 1 | < \epsilon\}$ where 
$d_Q(f_1,f_2) = \inf\{\epsilon > 0: Q(\{x:|f_1(x)-f_2(x)| > \epsilon\}) < \epsilon \}.$ Under
the assumption that the  prior over $\sigma$ assigns positive probability to every $\epsilon-$neighborhood
of $\sigma_0,$ one has: 

\begin{thm} 
    Let $f^{LX}$ be a randomly generated curve from the prior defined in $\S 2{\cdot}2$ 
         with $f_0\in \mathcal{F}^{H-1}.$ If $P_{f_0,\sigma_0}$ is the joint distribution of $\{y_i\}_{i=1}^{\infty}$
         conditionally on $\{ x_i \}_{i=1}^{\infty}$,   
	$\{\mathcal{Z}_i\}_{i=1}^{\infty}$ is a sequence of open subsets in $\mathcal{F}^{H-1}$ that is defined by 
       $W_{\epsilon,n}$ for fixed designs or by $U_\epsilon$ for random designs, and $\Pi_n$ 
	is the posterior distribution of $f_0$ given 
	$\{y_i\}_{i=1}^{n},$ then 
	\begin{align*}
		\Pi_n(f \in \mathcal{Z}_{n}^{C}|y_1,\ldots,y_n) \rightarrow 0 \hspace{10mm} \textit{almost surely} \hspace{2 mm}  [P_{f_0,\sigma_0}].
	\end{align*}
		Further, for all $H$ odd if  $M < 0,$  this relation holds for $f_0 \in \mathcal{F}^{H+},$
		otherwise it holds for  $f_0 \in \mathcal{F}^{H-}.$ Similarly, for $H$ even if 
		$M  > 0,$ then  $f_0 \in \mathcal{F}^{H+},$ otherwise it holds for $f_0 \in \mathcal{F}^{H-}.$ 
\end{thm}

The proof of this consistency result follows from 
\citet{choi2007} and the prior positivity result above.  The condition on the
prior over $\sigma^{2}$  can be satisfied with an inverse-Gamma prior.

\subsection{Bayes Factors for Testing Curve Shapes}

A key feature of our approach is that it allows one to explicitly define the shape
of the curve through the $\alpha$ vector and place prior probability on
a class of functions having a given shape.  We use the term shape to correspond
only to the number and type of extrema in $\mathcal{X},$ which
is parametrized through $\alpha.$ When there are flat regions of $f_0$
the shape of the curve is not uniquely identifiable based upon the configuration of 
the $\alpha,$ and all hypothesis tests may be inconclusive. For
an example of this, see the consistency arguments for monotone curve testing in \citet{scott2015}. In what follows, we assume that $|f_0'(x)| > 0$ 
at all points in $\mathcal{X}$ except at 
the extrema to rule out the consideration of flat regions. 

\textit{Remark} As posterior consistency is guaranteed when there are flat regions, the assumption that $|f_0'(x)| > 0$ is not required for model fitting.

Let $\mathbb{H}_1$ and $\mathbb{H}_2$ denote two distinct and non-nested sets of $\alpha$ values, corresponding to distinct shapes.  These sets are defined by the number of $\alpha_h \in \mathcal{X}$, the number of
$\alpha_h \leq \inf(\mathcal{X}),$ and the number of $\alpha_h \geq \sup(\mathcal{X}).$  
One can compute $\mbox{pr}(Y | f_0 \in \mathbb{H}_1 )$ and $\mbox{pr}(Y| f_0 \in \mathbb{H}_2 ),$  with the corresponding Bayes factor between the two shapes being 
\begin{align}
 BF_{12} = 	\frac{\mbox{pr}(Y|f_0 \in \mathbb{H}_1)}{\mbox{pr}(Y|f_0 \in \mathbb{H}_2)}. \label{bayesFac}
\end{align}
This quantity is not available analytically, but can be estimated through posterior simulation 
by monitoring the $\alpha$ and $\beta$ vectors.  

Any two possible shapes falling within $\mathcal{F}^{H}$ can be compared using this approach.
Alternatively,  one may be interested in the hypothesis that $f_0$ is in a class of functions with at least $K$ extrema.  For example, one may wish to assess whether or not the function is monotone. In this case, one can define $\mathbb{H}_1$ to correspond to functions in $\mathcal{F}^H$ with $F$ or more extrema and $\mathbb{H}_2 = \mathbb{H}_1^c$ to functions with less than $F$ extrema.  The value of $H$ can be elicited as an upper bound on the number of extrema to avoid highly irregular functions.  For such tests, the following result holds.

\begin{prop} \label{prop1}
	Let $\mathbb{H}_1$ be the class of functions in $\mathcal{F}^{H}$ with $F$ or more extrema and $\mathbb{H}_2 = 
\mathbb{H}_1^c \bigcap \mathcal{F}^H$.  If $f_0 \in \mathbb{H}_1,$ then
	\begin{align*}
			B_{12} \rightarrow \infty 
	\end{align*}
	as $n \rightarrow \infty.$
\end{prop}

This result is a direct application of Theorem 1 in \citet{walker2004}. 
It follows from the fact that local extrema spline representations
having fewer than $F$ change  points can never be arbitrarily close to the 
function of interest, and, consequently, $\mathbb{H}_1$ will be supported given more data. 

\section{Posterior Computation}

We rely on \citet{godsill2001} to develop a reversible jump 
Markov chain Monte Carlo algorithm to sample between models. 
Consider moves between models 
$\mathcal{M}$ and $\mathcal{M'},$ where the model $\mathcal{M'}$ has one extra
knot that is a child of a node also in $\mathcal{M}.$ 
As described further in the supplemental material,  most of the local 
extrema spline basis functions for model $\mathcal{M}$ and $\mathcal{M'}$ are 
identical, with only $j+2$ functions being different. 
Let $\beta_{-\mathcal{M}}$ denote the coefficients on all the splines that are the same
as well as $\sigma^{2}, \pi$ and $\lambda$, which are parameters shared between both models.  
The remaining spline coefficients are $\beta_{\mathcal{M}}$ and $\beta_{\mathcal{M}'}$ 
for models $\mathcal{M}$ and $\mathcal{M'}$, respectively.  
As in Godsill, given the shared vector $\beta_{-\mathcal{M}}$, we  
marginalize  $\beta_{\mathcal{M}}$ and $\beta_{\mathcal{M'}}$ out of the posterior
to compute $p(\mathcal{M}'|Y,\beta_{-\mathcal{M}})$ and $p(\mathcal{M}|Y,\beta_{-\mathcal{M}}).$
This marginalization requires numerical integration of multivariate normal distributions, which
are performed using \citet{genz1992} and \citet{genz2000}.
The probability of a move between two models is determined by the ratio
\begin{align}
		h &= \frac{q(\mathcal{M};\mathcal{M}')p(\mathcal{M}'|Y,\beta_{-\mathcal{M}})}{q(\mathcal{M}';\mathcal{M})p(\mathcal{M}|Y,\beta_{-\mathcal{M}})}, \label{metrat}
\end{align}
where a knot insertion is made with probability $\min(1,h),$ a knot deletion is made with probability $\min(1,1/h),$
and $q(\mathcal{M};\mathcal{M}')$ is the transition probability between $\mathcal{M}$ and $\mathcal{M}'.$

All proposals are made between  models that are nested and differ by only one knot.  When the current model has no children we propose a knot insertion with probability $1.$ Otherwise, the proposal adds or deletes a knot with probability $1/2,$ and the inserted or deleted knot is chosen with uniform probability.  For a knot insertion, that is, as we are going from model $\mathcal{M}$ to $\mathcal{M'},$
the available knots are represented by all failures in the branching process that generated $\mathcal{M}$. 
For a knot deletion, that is one goes from model $\mathcal{M'}$ to $\mathcal{M},$ 
this represents all of the nodes in the branching process that generated $\mathcal{M'}$ 
that do not have any children.  All other parameters, including the spline coefficients,
are sampled in Gibbs steps described in the supplement.

The posterior distribution is often multimodal, with the above sampler often getting stuck in a single mode.
This occurs when widely different parameter values have relatively large support by the data, 
with low posterior density between these isolate modes.  To increase the  probability  of jumps 
between modes, a parallel tempering algorithm \citep{geyer1991,geyer2011} is implemented, which is 
fully described in the supplemental material.

\section{Simulation}
We investigate our approach through simulations for functions
having $0,1,$ or $2$ local extrema interior to $\mathcal{X}.$ 
For all simulations, we place a $Ga(1,1)$ prior over  $\sigma.$ 
For the hyper prior on $\pi$, we let $\nu = 2$ and $\omega = 18$
which puts a low probability of favoring flat curves. Additionally, for the hyper prior over $\lambda$,
we let $\delta = 0{\cdot}2$ and $\kappa = 2,$ which favors smaller values of $\beta.$
Finally, all local extrema splines were constructed using B-splines of order $2$ with $M = 100.$

The Markov chain Monte Carlo algorithm was implemented in the R programming 
language with some subroutines written in C++ and is 
available from the first author.  Depending on the complexity of the 
function being fit, the algorithm took between $60$ and $90$ seconds per $50,000$ samples
using one core of a $3{\cdot}3$ gigahertz Intel i7-5830k processor. 
Parallelizing the tempering algorithm on multiple cores may substantially
reduce the computation time.  Additional information
on the convergence of the algorithm, as well as impact of the B-spline
order used, is examined in the supplemental material. 

\subsection{Curve Fitting}
We compare the local extrema spline approach to other 
nonparametric methods, including Bayesian P-splines  \cite{lang2004}, 
a smoothing spline method described in \citet{green1993},
and a frequentist Gaussian process approach described in chapter $5$ of \citet{shi2011}.  
We consider seven different curves each having between $0$ and $2$
extrema, and compare the fits of the other approaches to a 
local extrema spline specified to have at most $H=2$ change points.  
The following true curves are  investigated
\begin{align*}
	\begin{array}{rlrl} 
	f_1(x) =& 10 x^{2}                         & f_2(x) =& 2+20\Phi\left\{(x-0{\cdot}5)/0{\cdot}071\right\}\\
	f_3(x) =& 5 \cos(\pi x)   & f_4(x) =& 10(x-0{\cdot}5)^2 \\
	f_5(x) =& -2{\cdot}5 + 10\exp\left\{-50(x-0{\cdot}35)^2\right\} &	 f_6(x) =& 1 + 2{\cdot}5 \sin\left\{2\pi(x+8)\right\}+10x \\
	f_7(x) =& 5 \sin(2\pi x)/(x+0{\cdot}75)^3 - 2{\cdot}5(x+10{\cdot}5)
	\end{array}
\end{align*}
We assume $y_i = f_j(x_i) + \epsilon_i$ with $\epsilon_i \sim \text{N}(0,\sigma^2).$
Functions $f_1,f_2$ and $f_3$ are monotone, functions $f_4$ and $f_5$ have one change point, and $f_6$ and $f_7$ have two change points. For each simulation, a total of $100$ equidistant points are sampled 
across $\mathcal{X}=[0,1].$  In the simulation, we consider two variance conditions $\sigma^2  = 4$ and
$\sigma^2 = 1.$ For each simulation condition, $250$ data sets were generated, fit and compared using
the mean squared error, $n^{-1} \sum^n_{i=1} (\hat{f}(x_i)-f(x_i))^2,$  for the local extrema spline, smoothing spline, Bayesian P-spline, and Gaussian process approaches. 

For the local  extrema approach, we collected $50,000$ Markov chain Monte Carlo samples, with the first  
$10,000$ samples disregarded as burn in. For the parallel tempering algorithm,
we specify $12$ parallel chains with $\{\kappa_1,\ldots,\kappa_{12}\} = \{1/30,1/24,1/12,1/9,1/5,1/3{\cdot}5,1/2,1/1{\cdot}7,1/1{\cdot}3,1/1{\cdot}2,1/1{\cdot}1,1\},$ and monitor the target chain with $\kappa_{12}= 1.$ The P-spline approach was defined using $30$ equally
spaced knots, and the prior over the second order random walk smoothing parameter was given 
a $\text{IG}(1$,0${\cdot}$0005) distribution, which  was one of the 
recommended choices in \citet{lang2004}. In this approach, 
$25,000$ posterior samples were taken disregarding
the first $5,000$ as burn in. For the smoothing spline method, the R
function `smooth.spline' was used. Finally, the Gaussian process approach 
used a frequentist implementation given in the R package `GPFDA.'  

Table  \ref{table1}  gives the integrated mean squared error
of the local extrema approach as compared to the  other approaches. 
All numbers marked with an asterisk are significantly different
from the local extrema approach. In all conditions, the local 
extrema approach has an integrated mean square error that is 
numerically less than the other approaches, and, in most of these
conditions the value is significantly different at the 0${\cdot}$05
level, indicating the local extrema approach was superior, and, in some cases,
this improvement resulted in integrated mean square errors that were $40\%$
less than the closest competing method.  Generally, when there is 
high signal to noise ratio the methods perform similarly. 
However, in regions where the signal to noise ratio
decreases, specifically in flat regions, the local extrema approach was superior as 
it removed artifactual bumps from the estimate.

\begin{table}
	\centering
	\caption{Integrated mean squared error for all functions. For each function, the top row
	represents the simulation condition $\sigma^2 = 4$ and the bottom row represents the 
	simulation condition $\sigma^2 = 1.$ Asterisks signify the number is significantly
	different than the Local Extrema spline at the 0${\cdot}$05 level. For display purposes, 
	all numbers are multiplied by 10. }
	\label{table1}

	\begin{tabular}{clllll}
						&    	        &	Local 	&	Smoothing       &Bayesian          &  Gaussian  \\
	True Function   	&     &   Extrema Splines &  Splines  &  P-Splines&              Process \\ [5pt]
\multirow{ 2}{*}{$f_1$}	&		        &	1$\cdot$60	&	2$\cdot$11$^*$	 &	2$\cdot$28$^*$	&	2$\cdot$15$^*$	\\
						&		        &	 0$\cdot$49	&	0$\cdot$58	&	0$\cdot$55	&	0$\cdot$71$^*$	\\ [3pt]
											
\multirow{ 2}{*}{$f_2$}	&		        &	2$\cdot$59	&	4$\cdot$19$^*$	& 3$\cdot$82$^*$		&	5$\cdot$26$^*$	\\
						&		        &	0$\cdot$09	&	0$\cdot$13$^*$	& 0$\cdot$11$^*$		&	0$\cdot$15$^*$	\\ [3pt]
											
\multirow{ 2}{*}{$f_3$}	&				&	 1$\cdot$57	&	2$\cdot$43$^*$	&	2$\cdot$26$^*$	&	2$\cdot$64$^*$	\\
						&				&	0$\cdot$49	&	0$\cdot$67$^*$	&	0$\cdot$92$^*$	&	0$\cdot$79$^*$	\\ [3pt]
											
\multirow{ 2}{*}{$f_4$}	&				&	1$\cdot$70   &	2$\cdot$10$^*$	& 2$\cdot$15$^*$		&  1$\cdot$90$^*$	\\
						&				&	0$\cdot$49	&	0$\cdot$56$^*$	& 0$\cdot$49		&  0$\cdot$59$^*$	\\ [3pt]
											
\multirow{ 2}{*}{$f_5$}	&				&	 2$\cdot$55	&	3$\cdot$69$^*$	&	3$\cdot$39$^*$	&	3$\cdot$90$^*$	\\
						&				&	0$\cdot$61	&	1$\cdot$12$^*$	&	0$\cdot$98$^*$	&	1$\cdot$14$^*$	 \\ [3pt]
																					
\multirow{ 2}{*}{$f_6$}	&				&	2$\cdot$17	&	2$\cdot$57	&	5$\cdot$16$^*$	&	2$\cdot$44	\\
						&				&	0$\cdot$69	&	0$\cdot$72	&	0$\cdot$72	&	0$\cdot$79$^*$	 \\ [3pt]
											
\multirow{ 2}{*}{$f_7$}	&				&	2$\cdot$38	&	3$\cdot$39$^*$	&	3$\cdot$96$^*$	&	3$\cdot$30$^*$	\\
						&				&	0$\cdot$66	&	1$\cdot$05$^*$	&	0$\cdot$85$^*$	&	0$\cdot$90$^*$	\\
					
	\end{tabular}
\end{table}

\subsection{Hypothesis Testing}
We perform a simulation experiment investigating the method's ability to
correctly identify the shape of the response function. This is done for three
sets of hypotheses. In the first case, the null hypothesis is the
set of non-monotone functions and the alternative, $\mathbb{H}_1$, is the set of all
monotone increasing functions. In the second test, the null consists of all 
monotone functions and the alternative, $\mathbb{H}_2$,
is all non-monotone functions. Finally, for the third test 
the null hypothesis is the set of functions having at most one change point, 
and the alternative, $\mathbb{H}_3,$ is the set of functions with two change points first having
a local maximum followed by a local minimum. Functions are defined on $\mathcal{X} \in [0,1].$
The nine functions used in this simulation are

\begin{align*}
	\begin{array}{rlrl} 
					\multicolumn{4}{c}{\mathbb{H}_1} \\
		g_1(x)& =2+0{\cdot}5x +\Phi\{(x-0{\cdot}5)/0{\cdot}071\}   & g_2(x) &=0{\cdot}5 \sin\{2\pi (x+8)\} + 4{\cdot}75x \\[3pt]
		g_3(x)& \multicolumn{3}{l}{= 1 + 2{\cdot}25 x} \\   [3pt]
						\multicolumn{4}{c}{\mathbb{H}_2} \\
		g_4(x)& =-2(x-0{\cdot}75)^2     & 	g_5(x) &= 1+2x-1{\cdot}56 \exp\{-50(x-0{\cdot}5)^{2}\} \\[3pt]
		g_6(x)& \multicolumn{3}{l}{  = 15(x-0{\cdot}5)^31_{(x<0{\cdot}5)} +0{\cdot}3(x-0{\cdot}5)-\exp\{-250(x-0{\cdot}25)^2\}} \\[3pt]
					\multicolumn{4}{c}{\mathbb{H}_3} \\
		g_7(x)& = 0.85\sin\{2\pi(x+8)\} + 4{\cdot}75x   & g_8(x)&= g_5(x)\\   [3pt]
		g_9(x)&\multicolumn{3}{l}{=5\sin(2\pi x)/(x+0{\cdot}75)^3 - 2{\cdot}5(x+10{\cdot}5) + 2}.\\[3pt] 
	\end{array}
\end{align*}

For the simulation, data are generated assuming $y_i = g_j(x)  + \epsilon_i$, where
$\epsilon_i \sim \text{N}(0,\sigma^2)$ and $\sigma^2=1.$ 
We consider sample sizes of $n = 100, 200, 300,$ and  $400,$ with
$50$ data sets constructed where points are sampled evenly across $\mathcal{X},$
for each sample condition. The local extrema approach
is specified as above except $150,000$ posterior samples are taken with the first $10,000$
disregarded as burn in.  For tests $\mathbb{H}_1$ and
$\mathbb{H}_2$, the local extrema approach
is compared against the Bayesian method of \citet{salomond2014} as well as 
the frequentist methods of \citet{baraud2005} and \cite{wang2011}.  For the method
of Baraud et al. we use the test where $\ell_n = 25,$ and for the method of Wang and Meyer
we use $k=4$ splines, which were the most powerful tests presented in the respective articles. 

The Bayesian tests produce Bayes factors, while the frequentist tests have corresponding test statistics. An important question is how to choose thresholds for concluding in favor of the null or alternative so that the tests are calibrated in the same manner.  We compare the methods based upon area under a receiver operating curve. This approach allows an objective comparison between the testing 
approaches.  For the simulation, the false positive rate was computed from the values of the
test statistics for the other functions not in the test set.  For example, when the functions in hypothesis 
$\mathbb{H}_1$ were considered, the test statistics for functions in hypotheses $\mathbb{H}_2$ and $\mathbb{H}_3$ were used. 

\begin{figure}
	\centering
		\includegraphics[width=0.8\textwidth]{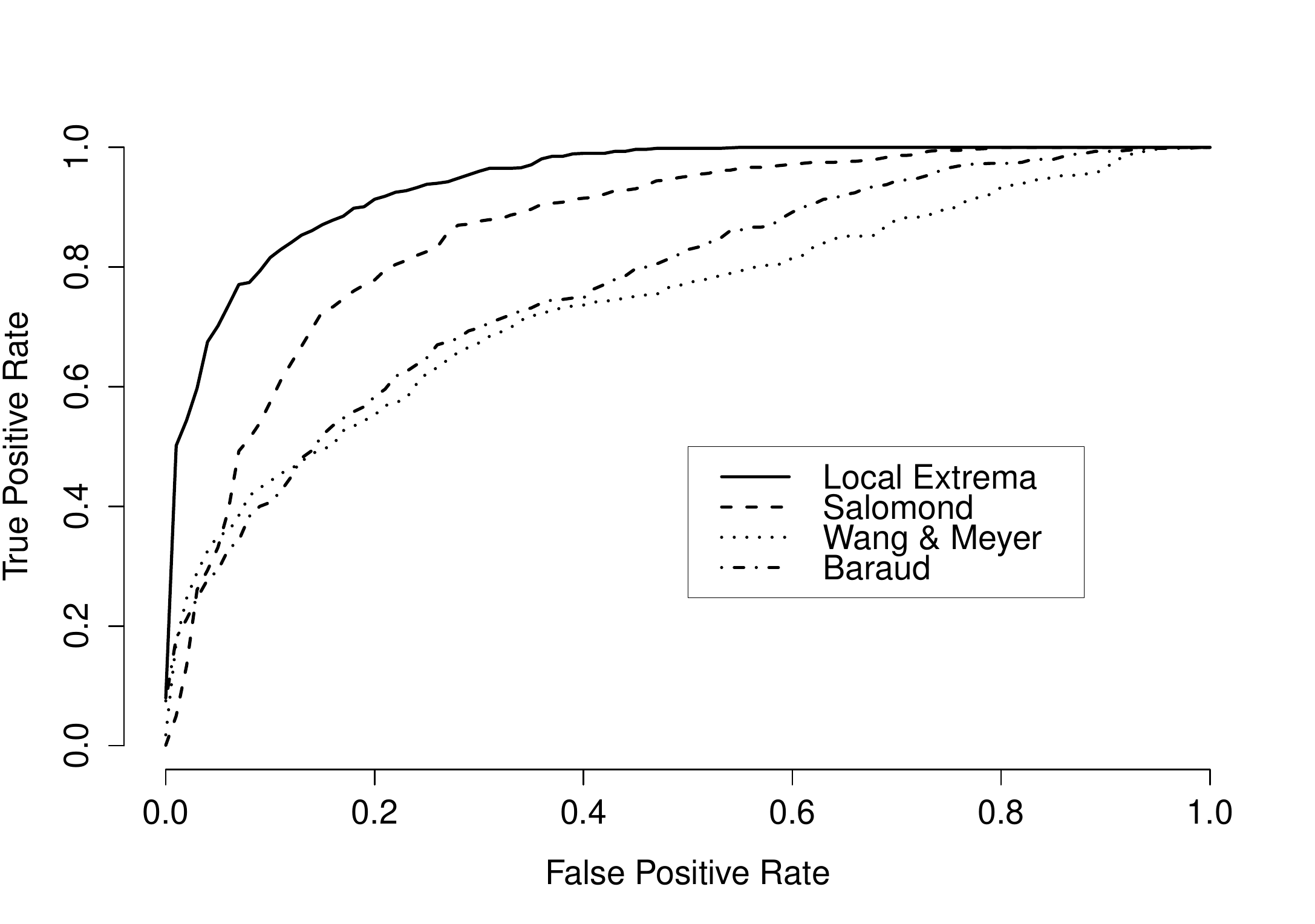}
	 \caption{The receiver operating curve for the four tests defined for hypothesis $\mathbb{H}_1$ for all $1,400$ simulations. }
\label{fig:ROC}
\end{figure}

Figure \ref{fig:ROC} shows the receiver operating curve for hypothesis $\mathbb{H}_1.$ 
This figure shows that the local extrema approach is superior to the other three approaches
across all false positive rates.  Further, the estimated area under the receiver operating
curve is $0{\cdot}94,$ which is excellent and better than the approach of Salomond at $0{\cdot}86$, 
Baraud at $0{\cdot}77,$ and Wang and Meyer at $0{\cdot}74.$ When looking at the impact of sample size on the tests, 
the power of the local extrema approach increases as the sample size increases, 
does so at a rate greater than competitors, and is similarly superior for hypothesis
$\mathbb{H}_2,$ data not shown.

For hypothesis $\mathbb{H}_3$ there is not an equivalent methodology in the literature, but performance of our 
approach is excellent. The area under the receiver operator curve
is $0{\cdot}937.$ For the Bayes factor cut point of $6$, table  \ref{table3} gives results across all simulation conditions. 
Our test achieves high power for function $g_7,$ even though this function is only slightly
different than $g_3$.  Function $g_8$ is the same as $g_5$,  and this simulation gives evidence that the departure from
monotonicity, which is concluded with high power hypothesis $\mathbb{H}_2$, may be due to the pronounced `U' shape in the data, and not necessarily due to the fact that there are two extrema. As evident by the observed power, 
this feature requires more data to conclude $\mathbb{H}_3$.

\begin{table}
	
	\centering
	\caption{Percent of samples where the model was correctly chosen as having 
	two extrema.}

	\label{table3}
	\begin{tabular}{ccccc}
Function		&	\multicolumn{4}{c}{n}							\\
		&	100	&	200	&	300	&	400	\\[5pt]
$g_7$		&	78	&	90	&	98	&	96	\\
$g_8$		&	14	&	32	&	22	&	46	\\
$g_9$		&	76	&	88	&	98	&	100	\\
\end{tabular}

\end{table}

\section{Applications}

\subsection{Estimating Muscle Force} 

When studying the ability of a muscle to adapt to exercise protocols, muscle force tracings are 
often used. One approach involves first activating the muscle and then after a short period of time 
moving the joint through the range of motion \citep{baker2008}. It is expected
that the muscle force quickly obtains a maximum force with the observed force decreasing
until joint movement; however, the observed force may plateau and not decrease before
movement. When the joint is moved, there is an expected increase in the force output until the joint
reaches a specific angle, after which, the observed force decreases until the joint reaches
its original position.  When the joint returns to its original position, the muscle remains activated
and the force output is non-increasing until deactivation. 
Estimation of this muscle force curve may allow better understanding of adaptation or maladaptation
following exercise, but it is important to include known biophysical constraints in curve estimation.

\begin{figure}
	\centering
	\includegraphics[width=0.8\textwidth]{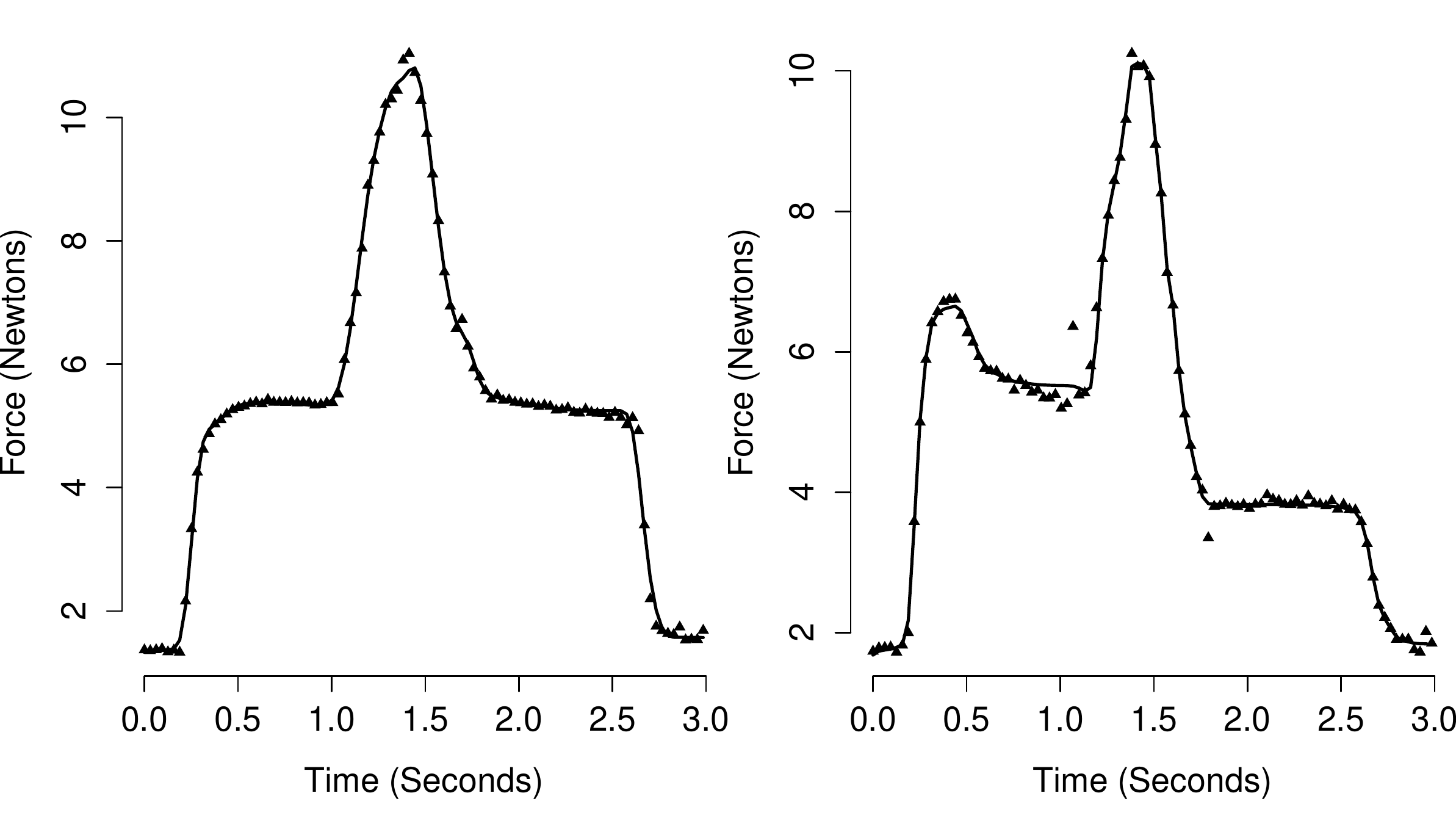}
		\caption{Fit of the local extrema spline,  black line, to observed muscle force data, solid triangles.}
		\label{fig:MFFig}
\end{figure}

We model two force tracings ($n=96$ per tracing) using a local extrema spline having at most $H=3$ local extrema.  Consistent with prior knowledge of a very high signal to noise ratio, we place a $\text{Ga}(2000,1)$prior on $\sigma^{-2}.$ We also applied frequentist smoothing splines, Gaussian processes, and Bayesian P-splines. Competing methods are close to interpolating the data points, leaving unwanted artifactual bumps in the function estimate.
However, as seen in Fig. \ref{fig:MFFig}, the local extrema spline obtains an estimate restricted to the
known shape and robust to minor local fluctuations. Further, when the force tracing exhibits a single maxima, as in the left plot, the local extrema spline can readily distinguish between this shape,  and
a shape which has two maximum, as in the right plot, with no change in the model.
 
\subsection{Seasonal Influenza and Pneumonia Death Rate}
	In temperate climates, the prevalence of influenza peaks in the 
	winter months while dropping in the warmer months. Estimating this seasonal effect
	as well as departures from this effect, may be of interest 
	when estimating the magnitude of an influenza epidemic. Here, we expect a peak in the 
	winter months followed by a trough in the summer months. Parametric models
	for this pattern may not be adequate to model the observed phenomena, and smoothing 
	approaches do not guarantee this pattern.  We use local extrema
	splines, setting $H = 2$, to estimate this trend for Virginia, North Carolina and South Carolina 
	for data collected by the Centers for Disease Control and Prevention National Center 
	for Health Statistics Mortality	surveillance branch.
	
\begin{figure}
	\centering
	\includegraphics[width=0.8\textwidth]{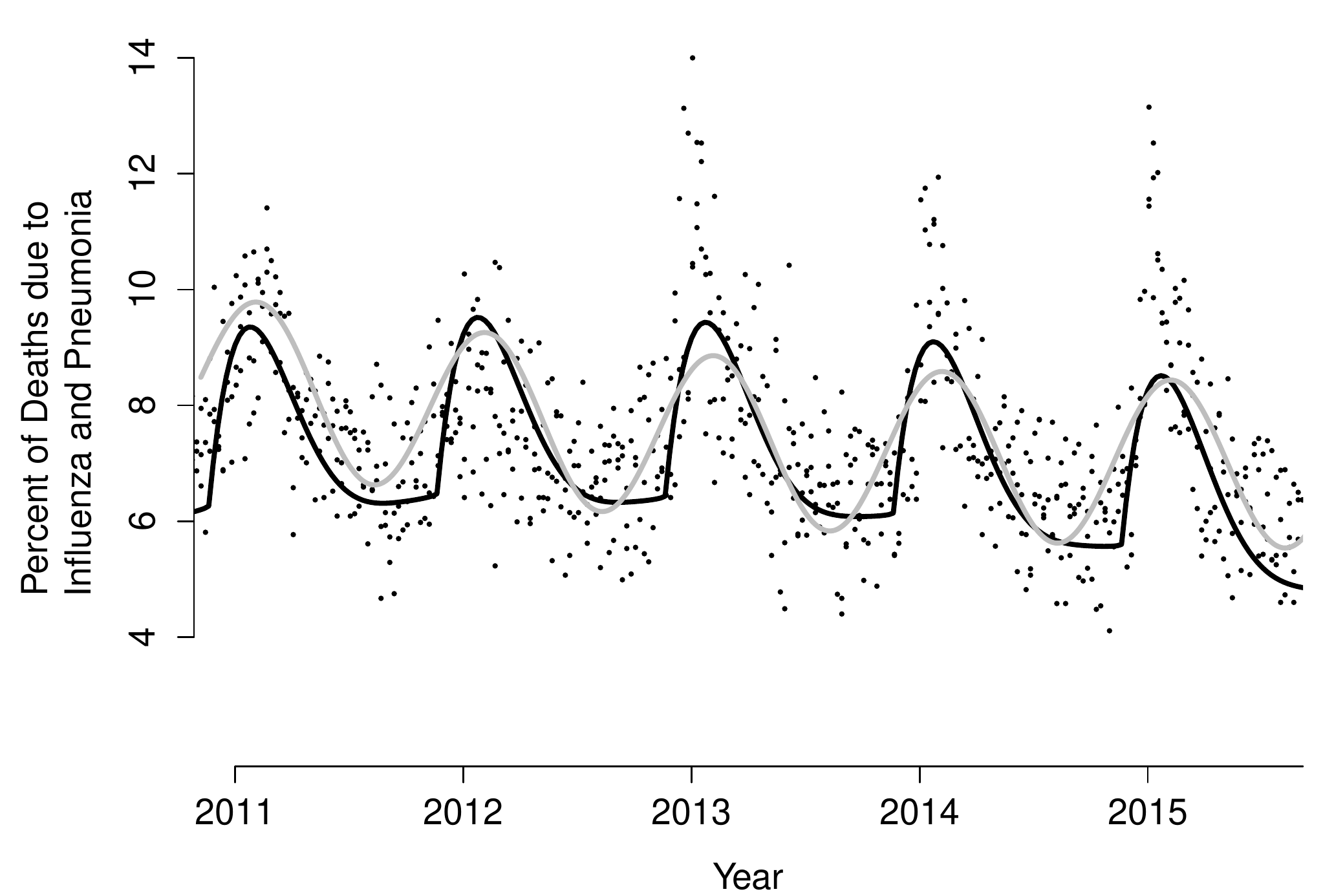}
		\caption{Estimate of the expected rate of seasonal influenza and pnuemonia deaths using the local
		extrema spline, black line, compared to the observed rate of influenza and pnuemonia deaths
		estimated using the Center for Disease Control's standard approach, gray line. }
		\label{fig:CFPFig}
\end{figure}

	Figure \ref{fig:CFPFig} plots the estimated mortality rates. These 
	rates are	estimated using an additive model defined by a 
	quadratic trend representing a decrease in mortality over time, a 
	seasonal component defined using local extrema spline, and a P-spline that represents
	departures from the overall trend. In this figure, the black curve represents the
	seasonally adjusted trend using the local extrema spline.   This seasonal component is different
	than the trend published by the Centers for Disease Control, gray line \citep{viboud2010}.
	The main difference between the two is the asymmetry in the local extrema 
	approach during the winter months, which can not not be captured by a single 
	sinusoidal function.

\section*{Acknowledgement}
This research was partially supported by a grant from the National Institute of 
Environmental Health Sciences of the United States National Institutes of Health.  The 
authors would like to thank Brent Baker for sharing the isometric muscle force
data.

\appendix

\section*{Appendix 1} 
\subsection*{Proofs of results}
\begin{proof}[of Proposition 1]
It is well known that $ \sum_{k=1}^{K+j-1} \beta_k B_{(j,k)}(x)$ is continuous for
$j \geq 1$ and for all $x \in \mathcal{X}$. Further, $\prod_{h=1}^H (x - \alpha_h)$ is a polynomial; 
therefore, $\prod_{h=1}^H (x - \alpha_h) \sum_{k=1}^{K+j-1} \beta_k B_{(j,k)}(x)$ is 
continuous with  anti-derivative $\sum_{k=0}^{K+j-1} \beta_k B^{\ast}_{(j,k)}(x).$ 

If $\beta_k \geq 0$ for all $k \geq 1,$ then 
$\sum_{k=0}^{K+j-1} \beta_k B_{(j,k)}(x) \geq 0$ for all $x \in \mathcal{X}$
and $f'(x) = \prod_{h=1}^H (x - \alpha_h) \sum_{k=1}^{K+j-1} \beta_k B_{(j,k)}(x)$ 
can only change sign when $x = \alpha_h.$
Thus, there are at most $H$ local extrema interior to $\mathcal{X},$ with $f \in \mathcal{F}^{H}$. 
\end{proof}

\begin{proof}[of Theorem 1]
Consider $f_0 \in \mathcal{F}^{H},$ where $f_0$ has
exactly $H$ change-points.  Functions with less than $H$ change points can be modeled
by removing the required change point parameters from $\mathcal{X}$ and continuing
with the proof below.

 Let $f^{BS}$ be a taut B-spline approximation of 
$f_0$ of order $j+1$ defined on the knot set $\mathcal{T}$ having exactly $H$ extrema
such that 
\begin{align*}
||f_0-f^{BS}||_{\infty} <  \Delta C. 
\end{align*}
Here $f^{BS}$ is defined on $\mathcal{T},$ where $\Delta = \max_k\text{ }|\tau_k - \tau_{k+j}| < 1.$
As $f_0$ and $f^{BS}$ are continuous and differentiable, we define $C$ such that $ \|f_0\|_\infty < C < \infty$ and $\|f^{BS}\| < C.$ 
The measurable set of taut spline functions $L^{\ast}_{f^{BS}}=\{f^{BS}: ||f_0-f^{BS}||_{\infty} <  \Delta C\}$ can
be shown  to exist \citep{deBoor2001} and we define a map $\mathcal{G}: L^{\ast}_{f^{BS}} \rightarrow L^{\ast}_{f^{LX}}$
where $L^{\ast}_{f^{LX}}$ a subset of all possible local extrema spline functions with $H$ change points. 
Consider
\begin{align}
	\|f^{BS}-f^{LX}\|_{\infty} &= \underset{x\in \mathcal{X}}{\sup} |f^{BS}(x)-f^{LX}(x)| \label{startp1}
\end{align}
and let $\beta_0 = f^{BS}(0).$  For the exactly $H$ extrema  $\alpha^{BS}_1 < \alpha^{BS}_2\ldots <\alpha^{BS}_H$ in $f^{BS}$
defined by the taut spline, set $\alpha_h=\alpha^{BS}_h.$
Additionally, if $f^{BS}(\alpha^{BS}_1)-f^{BS}(0) \geq 0$ with $H$ odd, then set $M = -1;$ otherwise set $M = 1.$
In the case where $f^{BS}(\alpha^{BS}_1)-f^{BS}(0) < 0$ with $H$ odd, then set $M = 1$ otherwise set $M=-1.$
   
Rewriting the RHS of (\ref{startp1}) in a form based upon the derivative we have
\begin{align*}
&\underset{x\in\mathcal{X}}{\sup} \left| \int_{-\infty}^{x} \sum_{k=1}^{K+j-1} \kappa_k B_{(j,k)}(\xi)- \beta_k G(\xi) B_{(j,k)}(\xi) d \xi	\right|, \\
&\leq    \sum_{k=1}^{K+j-1} \underset{ x \in \mathcal{X}}{\sup} \left|\int_{\tau_k}^{x}\kappa_k B_{(j,k)}(\xi)- \beta_k G(\xi) B_{(j,k)}(\xi) d \xi	\right|, 
\end{align*}
where the derivative of $f^{BX}$ is based upon the derivative formula for B-Splines \citep{deBoor2001}
and $G(\xi) = \prod_{h=1}^H (\xi - \alpha_h).$ 

Because of the taut spline construction of $f^{BS},$ we know that for
all $k,h$ such that $\alpha_h \notin [\tau_{k},\tau_{k+j-1}]$ one 
has $sgn(\kappa_k) = sgn(G(x)),$ for all $x \in [\tau_{k},\tau_{k+j-1}].$ 
Here $sgn({\cdot})$ is the signum function.
On each of these intervals let 
\begin{align*}
\beta_k = \frac{\int_{\tau_k}^{\tau_{k+j-1}} \kappa_k B_{(j,k)}(\xi) d\xi}{ \int_{\tau_k}^{\tau_{k+j-1}}  G(\xi) B_{(j,k)}(\xi) d\xi}.
\end{align*}
As $B_{(j,k)}(x) \geq 0$, we have $\beta_k \geq 0;$  further, one has 
\begin{align*}
\int_{\tau_k}^{\tau_{k+j-1}} \kappa_k B_{(j,k)}(\xi) -\beta_k G(\xi) B_{(j,k)}(\xi) d\xi = 0
\end{align*}
for all intervals such that $\alpha_h \in [\tau_{k},\tau_{k+j-1}].$

For  the at most $H$ coefficients defined on splines that are nonzero in the intervals
$\alpha_h \in [\tau_{k},\tau_{k+j-1}],$ set these coefficients to zero. 
As there are a finite number of intervals whose error is non-zero
and $f^{BS}$  is bounded, 
the maximum error is at most $(H+1) (j+1)  \Delta C$ for any $x$ and 
\begin{align*} 
	\| f^{BX}-f^{LX}\|_{\infty} \leq (H+1)  (j+1)  \Delta C.
\end{align*}
Consequently, for any $\epsilon$, consider taut B-spline constructions on knot sets $\mathcal{T}$
such that $\Delta \leq \epsilon[\{2(H+1)(j+1)\}C]^{-1}$
that also have $\|f_0-f^{BS}\|_{\infty} < \frac{\epsilon}{2}.$
Then one has
\begin{align*}
\|f_0-f^{LX}\|_{\infty} &\leq \|f_0-f^{BS}\|_{\infty} +  \| f^{BX}-f^{LX}\|_{\infty}  = \frac{\epsilon}{2} + \frac{\epsilon}{2} = \epsilon\\
\end{align*}
\end{proof}

\begin{proof}[of Lemma 1]
The function $\mathcal{G}$ in Theorem 1 is measurable. Given $L^{\ast}_{f^{BS}}$ is measurable on some abstract measure
space one has $\mbox{pr}(\|f^{LX} - f_0\|_{\infty} < \epsilon |\mathcal{T}_{\mathcal{M}}) > 0$ for any $\epsilon > 0$ and 
some   $\mathcal{T}_{\mathcal{M}}.$ 
Given that the prior puts  probability over knot sets having knot spacings that are arbitrarily close, that is  
 $\Delta \leq \epsilon[\{2(H+1)(j+1)\}C]^{-1}$ as in Theorem 1, we 
conclude that $\mbox{pr}(\|f_0-f^{LX}\|_{\infty} < \epsilon) = \mbox{pr}(\|f^{LX} - f_0\|_{\infty} < \epsilon |\mathcal{T}_{\mathcal{M}})\mbox{pr}(\mathcal{T}_{\mathcal{M}}) > 0$ for all $\epsilon > 0$.\
\end{proof}

\begin{proof}[of Theorem 2]
We verify the conditions given in A1 and A2 of Theorem 1 
of \citet{choi2007}.
Given that there is positive prior probability (Lemma 1) within all neighborhoods
of $(f_0,\sigma^2),$ one can use \citet{choi2007}, section 4, to show the conditions of A1 of 
Theorem 1 are met. To verify A2 we have  that $\mathcal{F}^{H+}$ and $\mathcal{F}^{H+}$
are subsets of all continuous differentiable functions on $\mathcal{X}$ which were considered
in \citet{choi2007}; consequently,
we appeal to Theorem $2$ and $3$ of \citet{choi2007} to construct suitable tests for 
both random and fixed designs using  $W_{\epsilon,n}$ and $U_{\epsilon}.$ We need only verify (\text{iii}) in part A2.

As in \citet{choi2007}, assume that $M_n = \mathcal{O}(n^\alpha)$ with ${1}/{2} < \alpha < 1.$ 
We show that $\mbox{pr}(\|f^{LX}(x)\|_{\infty} > M_n) \leq C_0 \exp(-n C_1)$
and $\mbox{pr}(\|f^{'LX}(x)\|_{\infty} > M_n) \leq C_2 \exp(-n C_3)$ for some $C_0,C_1,C_2,C_3 > 0.$ 
Define $B^{\ast}_{(j,k,\mathcal{M},\alpha)}(X)$ as the design matrix given model $\mathcal{M}$
and a particular $\alpha$ configuration. 
Let $A = \underset{\forall \mathcal{M},k,\alpha,x}{\sup} |B^{\ast}_{(j,k,\mathcal{M},\alpha)}(X)|$ 
and $K_\mathcal{M}$ be the number of spline coefficients in model $\mathcal{M}$ then 
\begin{align*}
 \mbox{pr}\left(  \|f^{LX}(x)\|_{\infty} > M_n\right) &= \int  \mbox{pr}\left(\|\sum_{k}^{K_\mathcal{M}} \beta_k B^{\ast}_{(j,k,\mathcal{M},\alpha)}(X)\|_{\infty} > M_n \bigg|  \mathcal{M}\right) d\alpha \hspace{1mm} d\mathcal{M} \hspace{1mm} d\pi \hspace{1mm} d\lambda \\
&\leq \int \mbox{pr}\left(\sum_{k}^{K_\mathcal{M}} \|\beta_k B^{\ast}_{(j,k,\mathcal{M},\alpha)}(X)\|_{\infty} > M_n \bigg|  \mathcal{M},\beta > 0\right )  d\alpha  \hspace{1mm} d\mathcal{M} \hspace{1mm} d\pi \hspace{1mm} d\lambda \\
&\leq \int \mbox{pr}\left(\sum_{k}^{K_\mathcal{M}} \beta_k A > M_n \bigg|  \mathcal{M},\beta > 0\right )  d\alpha \hspace{1mm} d\mathcal{M} \hspace{1mm} d\pi \hspace{1mm} d\lambda 
\end{align*} 
and by the Chernoff bounds
\begin{align*}
&\leq \exp(-M_n t) \int \ \sum^{\mathcal{M}} \bigg\{ \bigg(\frac{\lambda-\pi t}{\lambda-t}\bigg)^{K_\mathcal{M}} pr\bigg( \mathcal{M}\bigg)\bigg\} d\alpha \hspace{1mm} \hspace{1mm} d\pi \hspace{1mm} d\lambda \\
\end{align*}
Now let $pr^{\ast}(\mathcal{M})$ be the probability of a branching process where $\zeta < 0.5$ is constant for all children, then
there exists a $\mathcal{K}$ such that $\{pr^{\ast}(\mathcal{M})\}^2 \geq pr(\mathcal{M})$ for all $\mathcal{M}$ such that
$K_\mathcal{M} \geq \mathcal{K}.$ Partition the sum into the finite sum where $K_\mathcal{M} < \mathcal{K}$ and the infinite
sum $K_\mathcal{M} \geq \mathcal{K}.$ As the finite sum is finite for all $0 < t < \lambda,$ one has
\begin{align*}
&\leq \exp(-M_n t) \int \ C_1 + \bigg[\sum^{K_\mathcal{M}\geq \mathcal{K}} \bigg(\frac{\lambda-\pi t}{\lambda-t}\bigg)^{K_\mathcal{M}}\bigg\{pr^{\ast}\bigg( \mathcal{M}\bigg)\bigg\}^2\bigg] d\alpha \hspace{1mm} \hspace{1mm} d\pi \hspace{1mm} d\lambda \\
&\leq \exp(-M_n t) \int \ C_1 + C_2\bigg[\sum^{K_\mathcal{M}\geq \mathcal{K}} \bigg(\frac{\lambda-\pi t}{\lambda-t}\zeta\bigg)^{K_\mathcal{M}}pr^{\ast}\bigg( \mathcal{M}\bigg)\bigg] d\alpha \hspace{1mm} \hspace{1mm} d\pi \hspace{1mm} d\lambda \\
&\leq \exp(-M_n t) \int \ (C_1 + C_2) d\alpha \hspace{1mm} d\pi \hspace{1mm} d\lambda 
\end{align*} 
where the last inequality exists as $\lambda$ is bounded above zero, which implies one can choose some $t < \lambda$ such that $\frac{\lambda-\pi t}{\lambda-t}\zeta < 1.$
This implies that
\begin{align*}
\mbox{pr}(\|f^{LX}(x)\|_{\infty} > M_n) &\leq C_0 \exp(-n C_1).
\end{align*}

A derivation similar to the above can be used to show the same holds for $\mbox{pr}(\|f^{'LX}(x)\|_{\infty} > M_n) \leq C_2 \exp(-n C_3).$
One can find a $B = \underset{\forall \mathcal{M},k,\alpha,x}{\sup} |B'^{\ast}_{(j,k,\mathcal{M},\alpha)}(X)|$ and substitute $B$ for $A$ and $B'^{\ast}_{(j,k,\mathcal{M},\alpha)}(X)$ for $B^{\ast}_{(j,k,\mathcal{M},\alpha)}(X)$ in the above derivation\qed.  
\end{proof}

\bibliographystyle{apalike}
\bibliography{biblo}

\end{document}